\documentclass{hotnets17}

\usepackage{times}  
\usepackage{enumitem}
\usepackage{subcaption}

\usepackage{xspace}
\usepackage{mathtools}
\newcommand{\name}{SensorInc\xspace}
\newcommand{\sssec}[1]{\vspace*{0.05in}\noindent \textbf{#1\xspace}}

\usepackage[
breaklinks=true,    
colorlinks=true,
]{hyperref}

\usepackage{breakurl}

\begin{document}



\title{Sensor as a Company: On Self-Sustaining
IoT Commons}



\author{
	Haojian Jin, Swarun Kumar, Jason I. Hong\\
	Carnegie Mellon University\\
    \{haojian@cs.,swarun@,jasonh@cs.\}cmu.edu
}

\maketitle

\begin{abstract}
Beyond the ``smart home'' and ``smart enterprise'', the Internet of Things (IoT) revolution is creating ``smart communities'', where shared IoT devices collectively benefit a large number of residents, for transportation, healthcare,  safety, and more.  However, large scale deployments of IoT-powered neighborhoods face two key socio-technical challenges: the significant upfront investment and the lack of information on local IoT needs. In this paper, we present \name, a new IoT deployment paradigm that incentivizes residents to design and manage sensor deployment through sensor \textit{liquefaction}. By turning shared sensors into liquid (i.e. tradeable) assets akin to a company stock or bond, users can design and invest in promising IoT deployments and receive monetary rewards afterward. We present the detailed design of \name and conduct two case studies (parking occupancy sensors and air pollution sensors) to study the self-sustainability and deployment challenges of such a paradigm.  

\end{abstract}

\section{Introduction}


While the Internet of Things (IoT) is starting to revolutionize our homes (``smart homes'') and work places (``smart enterprises''), perhaps their biggest potential is in enabling ``smart communities''. Imagine shared IoT sensors deployed through a neighborhood to collectively benefit its residents for transportation (e.g. smart parking~\cite{January266:online}), healthcare (e.g. noise~\cite{segura2016spatial} and air pollution~\cite{Hsu:2019:SPC:3301275.3302293} sensing), safety (e.g. IoT-based crime watch~\cite{baig2017future}), education (e.g. IoT experimental testbeds~\cite{putjorn2018investigating}) and more. Beyond the traditional technological challenges, this paper focuses on two major socio-technical challenges in making community-driven IoT self-sustaining and truly ubiquitous. First, deploying and managing a shared IoT network is expensive, requiring a significant upfront investment. For example, San Francisco's SFPark~\footnote{http://sfpark.org}, a project that deployed over 5000 parking occupancy sensors throughout the city center, required an installation cost of \$330 for each sensor, and a running cost of \$10 per sensor per month~\cite{zakharenko2019economics}. Second, the self-sustaining deployment of an IoT sensor network requires a deep understanding of the local environment and users' potential needs. The traditional approach to do so relies on pilot deployments~\cite{disalvo2017fruit,lee2015internet} or surveys~\cite{apthorpe2018discovering}. For instance, SFPark's 141-page document~\cite{January266:online} summarizes their sensor placement decisions through 30+ independent surveys over 3 years. Globally, 70\% of community IoT projects are stuck in so-called ``pilot purgatory''~\cite{mckinseyreport, lee2015internet} as deployers lack the information needed to risk a full-scale deployment.

Our key observations on this problem are two-fold. On one side, residents, who are also the potential users of IoT applications, have the best knowledge of potential IoT deployment opportunities. On the other side, IoT applications can be profitable, but they require careful design and huge upfront investment. Based on these two observations, we propose a new IoT deployment paradigm, \name, which aims to \textit{incentivize} local residents to \textit{design} and \textit{manage} sensor deployment in shared spaces through sensor \textit{liquefaction}.

This paper explores the feasibility of making community sensors a liquid asset. 
In finance, a liquid asset, such as a stock, bond, or mutual fund, is an asset that can be readily converted to cash. To be considered liquid, an asset must be in an established market, with a large number of interested buyers, and with the ability to transfer ownership easily~\cite{WhatInve12:online}. 
Community sensors are traditionally seen  not as liquid assets but more as common pool resources~\cite{ostrom1990governing}. 
For example, SFPark is mainly funded by Federal funding through the Department of Transportation's Urban Partnership Program, and the services enabled by these sensors are free to users.

\begin{figure}
	\includegraphics[width=\linewidth]{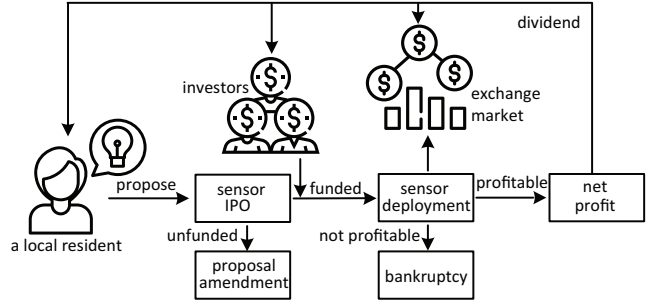}
	\caption{The lifecycle of a sensor micro company.}
	\label{fig:lifecycle}
\end{figure}

We argue that sensors are entitled to receive compensation when another service profits from the data produced by the sensor. These compensations will become the revenue of the sensor, which can be used to pay the bill for electricity, network, computation, storage, insurance, and potential human labor (e.g., maintenance cost). If the revenue can exceed the expenses\footnote{For simplicity, we do not discuss tax issues here.}, all the shareholders of the sensor will share the income (i.e., net profit) in the form of a dividend. To some extent, each sensor is a self-governed micro company -- analogous to a traditional company but with much lower valuation and overhead costs.

\begin{figure}
	\includegraphics[width=\linewidth]{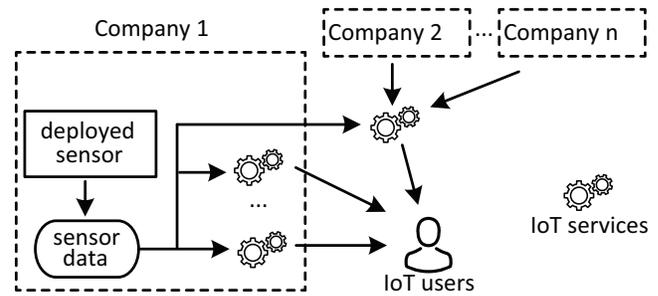}
	\caption{A sensor company generate revenues by either: 1) serving and being paid by end users directly; 2) providing data to and being paid by an external service, which may aggregate multiple sensors' data.}
	\label{fig:services}
\end{figure}

\name's market ecosystem (Figure~\ref{fig:lifecycle})  allows local residents to participate in various ways: 
\begin{itemize}[noitemsep,topsep=0pt]
	\item \textit{She can propose or invest in an Initial Public Offering for a new sensor}. An early technology adopter often perceives many IoT needs in their life~\cite{disalvo2017fruit}, e.g., knowing if there is parking space outside her favorite restaurant. Installing a sensor only for her is not cost-efficient, considering she may only use it occasionally. However, she anticipates others may share the same needs, so she can propose a sensor deployment plan, start a sensor micro company by selling the initial equities to interested investors (e.g., their neighbors).  
	\item \textit{As a shareholder, she can obtain a dividend from a sensor.} The sensor generates revenue by providing data to various services (Figure~\ref{fig:services}). The cash will flow from the end users to the service providers, and eventually to the sensor company. The company will redistribute the net profit to the shareholders in the form of a dividend.  
	\item \textit{She can trade the equity of a sensor in a public exchange market.} Each sensor company will have a real-time valuation based on the company value and the supply-demand relationship in the exchange market. The sensor value depends on the value of services it can offer. If the sensor company attracts more subscribers or develops new ways to use its data, the sensor value may appreciate over time. 
\end{itemize}

The rest of this paper outlines the technical framework in designing a self-sustainable IoT community infrastructure through sensor liquefaction (\S\ref{sec:core}) as well as the associated economic and social challenges. On the economic front, we study how best to effectively reward different classes of investors and stakeholders while minimizing risk. We present the design of \name\ IPOs and develop pricing and reward models for investors (\S\ref{sec:economic}). On the social front, we explore how day-to-day operations such as maintenance, accounting and legal services can be cost-effectively provided for low-value sensor micro companies. We present mechanisms that borrow from virtualization, a familiar computer systems approach, to aggregate several sensor micro-companies into larger virtual entities that can afford to pay for these services (\S\ref{sec:social}). We also discuss our thoughts and open questions pertaining to regulation, abuse, data ownership and privacy implications of our approach. Finally, we present a preliminary exploratory study targeting two applications  -- parking occupancy and air pollution monitoring (\S\ref{sec:casestudies}). 






\section{Sensor as a company}\label{sec:core}

At the core of \name is the design that each sensor is an independent, self-sustainable, and community-owned  "company"-like entity, which pays its shareholders a dividend regularly out of its profits. 
This notion of the sensor company brings several unique advantages to the  ecosystem. First, it separates the liabilities and obligations incurred by sensor operations from being the responsibility of the owners, releasing the proposer from a substantial and non-lasting commitment. Second, it connects investors, especially those who are interested in improving their local community, to a new type of liquid asset, providing the initial funding to deploy an IoT sensor network. 

\vspace*{0.05in}\noindent \textbf{Ecosystem: } The architecture of the sensor company ecosystem is analogous existing exchanges for other liquid assets such as stocks. However, we re-design this ecosystem to minimize paperwork and overhead in traditional stock exchanges. \name\ exchanges allow for IPOs, list companies and regularly updated valuations for all sensor companies as well as record associated ownership and financial transactions. Given the high volume and low value of individual transactions, we rely on existing micropayment frameworks~\cite{boddupalli2003payment,nakamoto2008bitcoin} (e.g. cryptocurrencies) to ease cost of trading. All \name\ exchanges are regulated to check abuse. 


\vspace*{0.05in}\noindent \textbf{\name-Specific Challenges: } Rather than exploring the technicalities  of an exchange or micropayments -- both well-studied financial systems~\cite{boddupalli2003payment}, this paper studies the economic and socio-technical challenges unique to sensor companies: 
\begin{itemize}[noitemsep,topsep=0pt]
	\item \textbf{Economic Interactions (\S\ref{sec:economic}): } How do we design the economic interactions among multiple parties (i.e., sensor companies, service providers, and investors) to build a self-sustainable ecosystem? Specifically, how should a sensor company charge for data/services, and reward its investors? How can policies be hard-coded to avoid the need for managerial roles (e.g. CEO / CFO)? 
	\item \textbf{Socio-Technical Implications (\S\ref{sec:social}): } How can a low-value sensor company afford the overhead of paying legal, accounting and maintenance services? How will traditional corporate administrative jobs need to evolve to fit the \name\ paradigm? How can \name\ be effectively regulated at scale to prevent abuse?
\end{itemize}

\section{Economic Interactions}\label{sec:economic}
This section describes the economic interactions between stakeholders in a \name\ company. Our key objective in pricing the economic interactions is ``self-sustainability''.


\sssec{Defining self-sustainability: } We say a \name\ company is self-sustaining if it benefits all the stakeholders, which incentivizes them to continue staying in the market, specifically: 

\begin{itemize}[noitemsep,topsep=0pt]
	\item The sensor company should be profitable.
	\item The investors should receive an investment return which is commensurate to their risk. 
	\item The price of the IoT services should fall within the end users' willingness-to-pay price amidst competition. 
	\item The proposer(s) should receive enough rewards for proposing high quality deployment plans. 
	\item All operational services (legal, accounting, technical maintenance, utilities, connectivity, etc.) should be paid for at fair market prices. 
\end{itemize}

The rest of this section discusses two goals of a self-sustaining \name\ company: (1) How do we profitably price services?; (2) How are profits shared with investors? 

\subsection{Valuing Services}
The intrinsic value of a sensor company comes from the services that a company can offer to its clients. Valuating such a service business largely relies on the expectation of future profit~\cite{Howtoval55:online}. The key challenge in doing so for a new \name\ company is the lack of prior profit history or comparables -- the  means of valuing traditional companies in the stock market today~\cite{Howtoval55:online}.

\name\ resolves this challenge by using the asset approach~\cite{Valuatio88:online} (i.e., the total cost to start the business) just past a sensor company's founding to value the company, due to the lack of a profit history. When the sensor company starts to receive a steady profit and distribute dividends, investors may be willing to pay more for the stock, the company valuation method will be shifted to the market approach~\cite{Valuatio88:online}.

A key unknown in company valuation is finding the right price for its services. Given that we are in the sensor context, one would traditionally price a sensor's value by the amount and quality of sensor data obtained~\cite{niyato2016smart}. Counter-intuitively, this does not lead to an accurate assessment of market price and is sub-optimal in terms of total revenue maximization. Instead, a sensor company should charge  clients based on the nature of the service (e.g., the marginal value of the service and the importance of the data to that service) rather than the amount and quality of the sensor data. The reason for this is that a data-hungry service does not necessarily provide more value than a service uses less data, since many contextual factors may impact users' willingness-to-pay price significantly.  If sensor companies charge data requests based on the data quantity, data hungry services are less likely to thrive, which will reduce the total revenue eventually. However, we emphasize that data quantity and quality will impact the profit margins of a sensor company, since it costs more resources to retrieve high accuracy data and transmit a large data file. 

\sssec{Deriving Initial \name\ Valuation: } Mathematically, we demonstrate our valuation method for a typical \name\ company as follows. Assuming a sensor company offers $K$ services to their users,  we use $\Phi = \{\phi_1, .., \phi_k\} $ and $ \Psi = \{\psi, .., \psi_k\}$ to denote the unit prices and the marginal cost of each service. Marginal cost is the additional cost of fulfill one more service/data request, e.g., the extra computation for analyzing, the network cost for communication. 

We assume that the number of service usages $n_i$ for $i$-th service is a monotonically decreasing function of the unit service price: 
\begin{align}\label{eq:price-usage}
n_i = f(\phi_i)
\end{align}
A lower unit service price $\phi_i$ will lead to more service usages $n_i$.
Then given that there are $N = n_1, n_2,.. n_K$ usages for each of the $K$ services,  we can formulate the monthly profit $P$ as follows: 
\begin{align} 
P = \left(\sum_{i=1}^{K} (\phi_i - \psi_i) * n_i - f_i \right) - F
\end{align}
where $F$ and $f_i$ are the fixed costs for the whole sensor company and the $i$-th service, respectively. Fixed costs are costs that do not change when the quantity of output changes. For example, the overall fixed cost $F$ includes the equipment depreciation, hardware maintenance, accounting services, etc. The $i$-th service fixed cost refers to the fixed cost to setup a new service, e.g., the IoT app program purchase fee.

\subsection{Return to Investors}

Having assessed a fair-market value for services and the sensor company's profits, \name\ companies need to choose whether to re-invest this profit in their own growth (akin to a growth stock, e.g. technology companies~\cite{Technolo44:online}) or pay back investors immediately. To minimize risk to investors, we choose the latter option which is similar in spirit to dividend stocks (e.g., utility companies in today's stock market). Such a fixed policy also obviates the need for managerial roles (e.g. CEO/CFO) to intervene with the financial model. A sensor company will distribute all the extra profits as dividends, after contributing to an internal reserve fund, to shareholders, rather than re-invest in itself to grow the business. The exchange market will be a transparent data-driven market, where each company will release a daily earnings report as well as be transparent about future outlook (which may be impacted by maintenance issues, sensor theft, etc.). 

Another important aspect of crafting dividends is to recognize two classes of investors: Pre-IPO investors and IPO investors. It is important to distribute higher rewards to investors who took a greater risk (i.e. the Pre-IPO investors).

\sssec{Deriving the basis for \name\ Dividends and Valuations:} 
We demonstrate an effective way to derive the market value and dividend and appreciation through IPO of a sensor company, using Price-to-Earnings ratio (P/E). We aim to design IPO sensor equities as a comparable investment to utility company stocks, which has an Annual Percentage Rate (APR) at around 4.1\%~\cite{The10Bes70:online}. Assuming the P/E of a sensor company is $x$, each individual stock price would be $s = x \times p$, where $p$ is the earning per share. Assuming all the profits are distributed as dividends, the APR for a single IPO sensor stock can be derived from P/E:
\begin{align} 
\alpha = \frac{p}{s} \approx \frac{p}{x \times p} = \frac{1}{x}
\end{align} 
Given an annual return of 4.1\%, the P/E for a sensor company would be 24.4 (average P/E for the utilities sector $\approx$ 23~\cite{Whatpric89:online}), and the value of a sensor company would be around 24 times of its yearly income.

Pre-IPO investors will receive a higher rewards commensurate with the risk they take. We assume pre-IPO investors achieve the reward mainly through sensor appreciation, when the company valuation method shifts from the asset approach to the market approach. However, the risk is that not all the sensor companies survive. Many of them may go bankrupt due to lack of sustainable users -- leading to a net loss to investors. Assuming that it takes $z$ months for a sensor starts to receive a steady revenue, and the initial pre-IPO value is $I$, the APR for a successful investment is: 
\begin{align}
    \alpha = (\frac{x \times P - I}{I})^\frac{12}{z} - 1
\end{align} 

We note that the above valuation method requires prior information such as the exact numbers of sensors.  Without this information, it is challenging to inform accurate rewards and the valuations. We feed empirical data into our model and discuss its feasibility in \S\ref{sec:casestudies}.

\section{Socio-Technical Implications}\label{sec:social}

This section studies two socio-technical objectives of \name: (1) fair and manageable staffing and service costs in a \name\ firm as well as ensuring payments for physical and network infrastructure; (2) regulating \name\ firms at scale to prevent abuse and ensure data privacy. 



\subsection{Staff and Services in \name}
\name\ companies are inherently of low-value, around hundreds of dollars at best. At these valuations, requiring each company to hire independent accounting, auditing, management, legal and daily operational and maintenance staff -- as is the case in the present corporate ecosystem -- would be unprofitable. On the flip side, existing trained personnel in accounting, legal, maintenance and other roles would simply find the fees paid by one sensor company too miniscule to offer their services at market  wages. 

\sssec{\name\ Virtualization: } \name\ addresses this challenge by relying on the virtualization, a familiar concept in computer networking and systems. Our approach specifically aggregates multiple individually small \name\ companies into one large virtual company~\cite{chun2003planetlab}, with a significantly larger valuation. This virtual entity can now afford to hire full time staff for administrative, management, legal and maintenance and accounting tasks at competitive rates. We note that the number and nature of staff needed for daily operations are statically specified during a \name\ company's IPO, negating the need for management roles (e.g. COO). \name\ performs aggregation at multiple levels: balance sheets, legal documents and financial transactions to ensure sufficient scale of the resulting entity. Virtualization can be directly offered as a service at the exchange with legal frameworks to ensure privacy and isolation between individual firms, and compliance with investor policies. 

An important challenge in virtualization is choosing portfolios of sensor companies that are aggregated. Choosing too many companies would aggregate firms that are otherwise not co-located making them difficult to manage. Choosing too few would mean they would not reach the critical mass to pay for staff and services at competitive wages. This means we would need to craft an optimization that virtualizes appropriately designed subsets of mutually-compatible \name\ entities.

We illustrate the mathematical formulation of such an optimization in a strawman scenario for brevity. Let the binary value $x_{ij}$ be equal to one if sensor company $i \in \{1, \dots, M\}$ is mapped to a virtual entity $j \in \{1, \dots, N\}$, where $N \ll M$. Let $V_i$ denote the valuation of sensor company $i$ and $C_j$ denote a compatibility score of sensor companies within the $j$th virtual firm, for e.g. based on the physical proximity of the sensors. Then we seek to maximize compatibility:
$$ \max \sum_j C_j $$ 
\vspace*{-0.2in}$$ \forall j, \sum_{ij} V_i x_{ij} \leq T \text{~~ ; ~~} \sum_i x_{ij} = 1, \text{~~ and ~~} \forall j, \sum_i x_{ij} =1$$
Where $T$ is a total valuation threshold. For certain definitions of $C$ (e.g. based on distance), the above can be approximately solved using well-known quadratic optimization algorithms~\cite{floudas1995quadratic} and relaxing $x_{ij}$'s  to reals between $0$ and $1$. 

\begin{figure}
	\includegraphics[width=\linewidth]{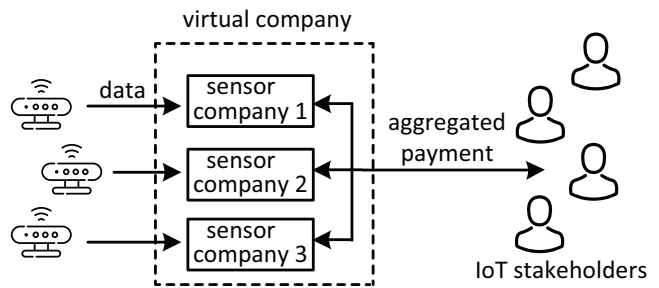}
	\caption{Sensor company virtualization.}
	\label{fig:sensorvirtualization}
\end{figure}

\sssec{Payments to Staff, Services and Utilities: } Sensor company virtualization can help aggregate payments made to various entities -- accountants, lawyers, managers and maintenance staff -- reducing the overhead costs and fees. Each virtual entity operates a single payment gate way to all staff and service providers. On the flip side, this will also spur an ecosystem of sensor company services offered by consulting firms that support micropayments for most mundane company tasks. We note that beyond staffing and services, aggregation also eases payments for physical and network infrastructure. We note that sensor companies may choose to aggregate their payments to land owners (companies, governments, universities, etc.), infrastructure and utility providers (network companies, electricity, etc.) to negotiate more favorable rates and reduce transaction fees.






\subsection{Regulation and Privacy}


\name\ exchanges must be carefully regulated to prevent abuse and minimize investor risk. 

\sssec{Regulation and Preventing Abuse: }To ensure high-quality and profitable firms listed on the exchange, we propose that sensor IPOs use an all-or-nothing funding model~\cite{Whyisfun51:online}, which means that the proposal will not get any investment unless the project reaches its funding goal. An undersubscribed IPO~\cite{Subscrib25:online} will  directly terminate the company. All-or-nothing funding, commonly used in the modern crowdfunding platforms, such as Kickstarter~\cite{Kickstar88:online} and Indiegogo~\cite{Crowdfun24:online},  helps cultivate the \name ecosystem in three ways: it reduces risks for investors, motivates the proposer to develop high-quality plans, and filters-out unqualified proposals.

\name\ must enforce careful reporting to prevent abuse. For instance, imagine insider-trading scenarios where managers sell-off all stock on a sensor company without reporting damage or theft of that sensor. \name\ exchanges must impose strict legal (e.g., fines) and institutional penalties (e.g. reputation scores, blacklisting, etc.) to check abuse. Given the scale of the \name\ ecosystem, the exchange must periodically monitor transactions for tell-tale patterns of abuse (e.g. managers selling-off stock in a coordinated way), akin to how this is done in the stock market today. \name\ must also facilitate seamless liquidation of assets in case of closure or bankruptcy of sensor companies.

\sssec{Data Ownership and Privacy: } Data is central to IoT, but who owns data collected by an IoT sensor? A panel at MIT Connected Things debated this question with no consensus on the answer~\cite{Thegreat1:online}. While we do not answer this difficult policy question in this paper, we argue that the \name\ ecosystem makes it easier to ensure compliance with any given data ownership and privacy policy. In particular, ownership of a sensor's data can be directly held by the \name\ company as opposed to any individual. Given that \name\ companies are listed in an exchange, this makes them easier to audit and regulate based on institutional or governmental privacy policies (e.g. GDPR~\cite{GeneralD36:online}). Contrast this with today's system -- where a privacy violation by a community sensor is hard to pin to one individual -- the deployer, the maintainer or the land owner, all of whom may deny accountability.

\section{Exploratory Study}\label{sec:casestudies}
We run an exploratory study, covering two application scenarios (air quality and parking occupancy) to understand the economic feasibility and constraints of \name. 


\sssec{Applications: } \label{sec:apps}
We study two different application scenarios that suit the \name\ paradigm: 


%

\noindent \textsc{(1) Air quality sensor: } A resident proposes to deploy an air quality sensor in her backyard. Different from traditional weather forecast systems, it provides air quality at the street/block level with more detailed information (e.g., particulate matter, carbon monoxide, ozone and PM 2.5).

{The cost of hardware and installation is \$179~\cite{Awair2nd65:online}. She will provide the electricity and network access to the sensor, as well as some necessary maintenance work (around 1 hour per month). She will also receive a \$10 hourly pay for the maintenance work and an incentive reward (5\% of the total revenue) if the sensor operates well in the past month.}~\footnote{Note all the numbers are not fixed. We expect IPO investors and proposers will reach their agreement through market equilibrium. }

{If the IPO is successful and the sensor company achieves a steady profit, the proposer will receive a 10\% company equity, through an Employee Stock Ownership Plan (ESOP). }

\begin{figure*}
	\centering 
\begin{minipage}{.20\textwidth} 
	\begin{tabular}{|p{3mm} |p{22mm} |}
		\hline \hline
		 & \small Service description  \\
		\hline
		\small \#1 & \small  Real-time air quality report   \\
		\hline
		\small \#2 & \small Weekly reports  \\
		\hline
		\small \#3 & \small 5-year air quality report (e.g., supporting real-estate investment) \\
		\hline
		\small \#4 & \small Least polluted routing feature on a mobile map app  \\
		\hline \hline
	\end{tabular}
	\caption{Services for air pollution sensors}
	\label{table:2}
\end{minipage}
~
\begin{minipage}{.22\textwidth}
	\centering \small
	\begin{tabular}{|p{3mm} |p{22mm} |}
		\hline \hline
		 & Service description  \\
		\hline
		\#1 & Real-time parking occupancy check \\
		\hline
		\#2 & Nearby street parking search \\
		\hline
		\#3 & Statistical report about the usages of the street parking spots outside of the restaurant, e.g., the occupancy rate and distribution.\\
		\hline \hline
	\end{tabular}
	\caption{Services for parking sensors.}
	\label{table:3}
\end{minipage}
~
\begin{minipage}{.22\textwidth}
	\small
	\begin{tabular}{|p{13mm}|p{8mm} |p{8mm}|}
		\hline \hline
		 & Air & Parking \\
		\hline
		IPO value & \$295 +200  & \$39 +200 \\
		\hline
		Income / yr & \$168 & \$23.75  \\
		\hline
		Market value & \$3904 & \$579.5 \\
		\hline
		\# of users & 43 & 29 \\
		\hline
		Return over pre-IPO & 788\% & 242\% \\
		\hline
		Proposer reward & \$390 & \$116 \\
		\hline
		Admin pay & \$10*12 & \$20*6 \\
		\hline \hline
	\end{tabular}
	\caption{\normalsize Financial metrics}
	\label{tab:met}
\end{minipage}
~~
\begin{minipage}{0.25\textwidth}
	\centering
	\includegraphics[width = \textwidth]{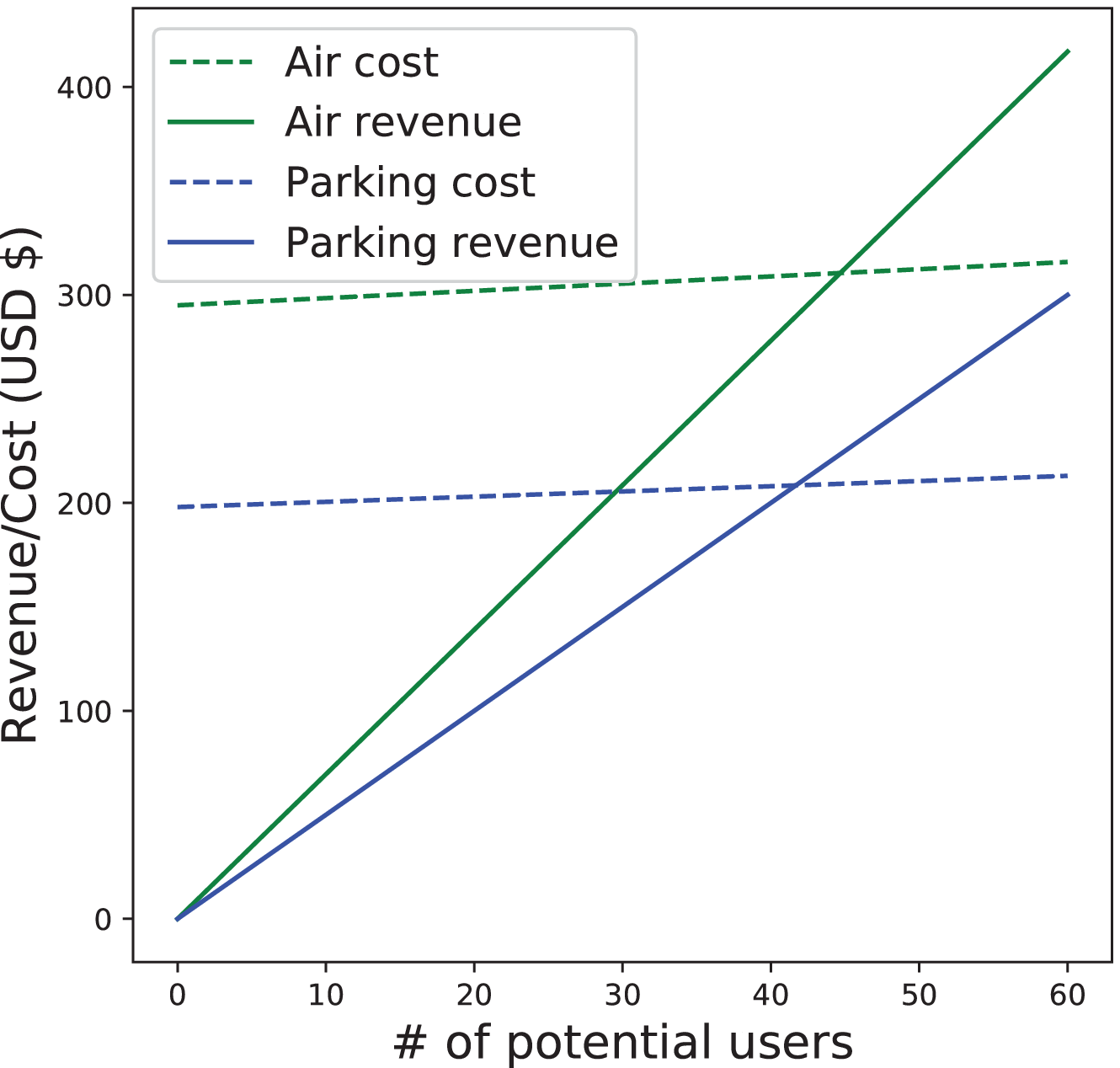}
	\caption{Expected cost and revenues}
	\label{fig:cost}
\end{minipage}
\begin{minipage}{0.35\textwidth}
~
\end{minipage}
\end{figure*}

\noindent \textsc{(2) Parking occupancy sensor:}
{A resident proposes to deploy parking occupancy sensors for the street parking spots next to her favorite restaurant. There are three street parking spots (\$2 per hour) next to the restaurant. If all these spots are occupied, a customer may either park at a paid parking building (\$8 minimum) or find another restaurant. }

{The proposal plans to install the sensors at the front of the restaurant with the approval of the restaurant owner. The cost of hardware and installation is \$39 per sensor~\cite{NestDete7:online}. The sensor company will hire a professional sensor management company to maintain the sensor, e.g., replacing batteries, checking network connection. The sensor company will charge a \$20 hourly pay, and receive an incentive reward (5\% of the total revenue) if the sensor operates well in the past month.}

{If the IPO is successful and the sensor company can achieve a steady profit, the proposer will receive a 20\% equities of the whole company, through the ESOP.}

\sssec{Pricing surveys: } \label{sec:apps}
As explained in \S\ref{sec:economic}, the intrinsic value of a sensor company comes from the services it can offer. To quantify the profitability, we first derived straightforward sensor usages from two sensor deployments (Table~\ref{table:2} and~\ref{table:3}), and then used marketing pricing surveys to determine the willingness-to-pay price for each service. Note, we did not exhaustively enumerate all the possibilities. Instead, we listed the common and mutually-exclusive services to illustrate economic feasibility and avoid double counting.

We first run a 100-participant Van Westedorp Pricing Model~\cite{lipovetsky2011pricing} survey to gauge consumers' perceptions of the willingness-to-pay price of each service independently. Through this survey, we found that users' usage frequencies highly correlates with the desired price. We then selected the common price points and recruited another 100 participants to study users' expected usage frequency regarding different prices: 
"\textit{If the price for each service usage is {[Free / \$0.01 / \$0.05... / \$5 / ...]}, how often would you expect to use such a service?}" We recruited participants through Amazon Mechanical Turk and each participant received a compensation between \$0.40 - \$0.8.  
Finally, we modeled the price-usage function (Eq.~\ref{eq:price-usage}) using the survey data by linearly  interpolating price points.

\sssec{Results: } \label{sec:apps}
Figure~\ref{fig:cost} illustrates the expected cost and revenues for sensor companies regarding the number of potential users in the first year. The two companies will break even if there are 29 users for the parking sensor and 43 users for the air quality sensors. We observe that our cost curves are nearly linear due to the relatively low overhead compared to traditional companies today. We assume that the hardware depreciation is 10\% of the purchase price per month and our survey respondents are representative. We also account the factor that many users are not ready to pay for IoT services: 99/350 of the responses indicate that they will not use such a service if it's not free. We select the service price which can maximize the revenue. The expected revenue per user is \$6.95 per year for air quality sensors, and \$5.0 for parking sensors. Table~\ref{tab:met} enumerates all the financial metrics of the two break even sensor companies, as well as the rewards for the stakeholders, at the end of the first operational year.

\section{Related Work}\label{sec:relatedwork}



\sssec{Crowdfunding}: Crowdfunding gathers funding for a project or venture by raising small amounts of money from a large number of people. For example,  Kickstarter~\cite{Kickstar88:online} and Indiegogo~\cite{Crowdfun24:online} help entrepreneurs pre-sell a product or service to launch a business concept. \name falls into another subcategory, equity crowdfunding which exchanges pledged money with shares of a company. However, \name\ is the first to explore this concept in IoT micro-funding context.  

\sssec{The Economics of Networks}: Studying the network incentives and pricing is an important topic in networking research. Past work has explored novel pricing/incentive framework for WiFi sharing~\cite{dimatteo2011cellular}, Cross-carrier data plan~\cite{zheng2017economic} and mobile ad hoc networks~\cite{buttyan2003stimulating,zhong2003sprite}. In contrast, \name seeks to incentivize IoT deployment, rather than networking alone.

\sssec{Public commons}: Shared IoT sensors are an emerging component of the future public commons. We apply many concepts and principles of commons in general (e.g., clear boundary, flexible execution~\cite{ostrom1990governing}) throughout \name's design.

\section{Conclusion and Open Questions}\label{sec:discussion}

This paper proposes \name, which transforms IoT sensors into liquid assets akin to a company stock, where users can design and invest in promising IoT deployments to receive monetary rewards. We present the detailed design of \name and conduct two case studies to study the socio-technical challenges of such a paradigm. We  believe that our work spurns broader open technical and policy questions on future community IoT infrastructure:

\sssec{Sensors vs. Actuators: } While this paper focused on IoT sensors, many other IoT devices such as actuators could also be treated as liquid assets in a similar spirit. However, two open questions remain when virtualizing actuators in particular. First, one would need to resolve conflicts where multiple customers seek to simultaneously operate an actuator. Second, ensuring privacy, regulation and safety of actuators in public spaces remains an open problem. 

\sssec{A Model for Self-Sustaining Testbeds: } Several infrastructure testbeds, e.g. those funded by NSF~\cite{NSF}, require a long-term sustainability plan -- a task challenging to ensure for heterogeneous IoT deployments. This paper potentially serves as a model to do so. A key open question remains: how can one accurately place a fair dollar price for the research utility of services and nodes rendered in a testbed?

\sssec{Future Jobs and Alternate Revenue Models: } While this paper assumes staff in an IoT company are akin to those in traditional companies, the new ecosystem may spurn a different type of contract-based job market. This market may create virtualizers-in-the-middle who aggregate multiple low-dollar value job tasks and assign them to contract labor. The simple and low volume nature of the tasks may also fuel the trend towards job automation~\cite{david2015there}. Workers may be paid in portfolios of sensor stock options as opposed to cash amounts. Our new \name\ ecosystem raises policy questions on worker rights, information ownership and private vs. public ownership of community IoT infrastructure. We emphasize the need for a broader and deeper study on the policy implications of our proposal.

\def\UrlBreaks{\do\/\do-}
\bibliographystyle{abbrv} 
\begin{small}
\bibliography{hotnets17}
\end{small}

\end{document}